\documentclass[a4paper,12pt,reqno,superscriptaddress]{revtex4}

\usepackage[centertags]{amsmath}
\usepackage{amsfonts}
\usepackage{amssymb}
\usepackage{amsthm}
\usepackage{newlfont}
\usepackage{stmaryrd}
\usepackage{mathrsfs}
\usepackage{euscript}
\usepackage{ulem}
\usepackage{color}

\theoremstyle{plain}

\theoremstyle{definition}

\theoremstyle{remark}

\numberwithin{equation}{section}

\newcommand{\opunit}{\text{1}\kern-0.22em\text{l}}

\DeclareMathAlphabet{\mathpzc}{OT1}{pzc}{m}{it}

\newcommand{\id}{\textrm{d}}

\begin{document}

\title{{\bf The modified Langevin description for probes \\ in a nonlinear medium}}

\author{Matthias Kr\"uger}
\email{mkrueger@is.mpg.de}
\affiliation{4th Institute for Theoretical Physics, Universit\"at Stuttgart, and Max Planck Institute for Intelligent Systems, Stuttgart, Germany}
\author{Christian Maes}
\affiliation{Instituut voor Theoretische Fysica, KU Leuven, Belgium}

\begin{abstract}
When the motion of a probe strongly disturbs the thermal equilibrium of the solvent or bath, the nonlinear response of the latter must enter the probe's effective evolution equation.  We derive  that induced stochastic dynamics using second order response around the bath thermal equilibrium.  We discuss the nature of the new term in the evolution equation which is not longer purely dissipative, and the appearance of a novel time-scale for the probe related to changes in the dynamical activity of the bath.  A major application for the obtained nonlinear generalized Langevin equation is in the study of colloid motion in a visco-elastic medium.
\end{abstract}
\maketitle

\section{Introduction}
When a probe is in contact with many degrees of freedom, one can often obtain a useful effective description by integrating out those bath degrees of freedom. Such techniques have a long history in both classical \cite{bha,vK,KO,zwanzig,zwa61,zwan,Dengler15} and quantum systems \cite{Caldeira81}, and  still constitute a very active domain of research. Probe motion is especially interesting -- and the theoretical treatment especially demanding -- in complex (e.g. visco-elastic or glassy) baths. Studies of those include use of two body equations valid in the dilute limit \cite{batchelor83,Beenakker84,Brady_Squires,Kruger}, density functional theory \cite{penna03b,rauscher07b}, mode coupling theory \cite{Gazuz08},  lattice models \cite{Benichou13}, and computer simulation \cite{Wilson, Winter}. Employing modern techniques such as optical tweezers, (active) probe dynamics can nowadays also be studied in precise experiments \cite{Wilson,Ruben14, beh}. 

Integrating out degrees of freedom of complex baths gives rise to a host of partly interconnected phenomena; Finite bath relaxation times lead to memory and possibly non-Gaussian nature of noise. Nonlinear terms, e.g. nonlinear in probe velocity, may arise, and render the noise dependent on the probe degrees of freedom (``multiplicative'' \cite{Risken}). 

In this manuscript, we develop a systematic expansion scheme for the effective Langevin description in media where nonlinear terms in probe displacement are important. This derivation is based on response theory, expanding the bath dynamics around the equilibrium state \cite{ub} where the probe is at rest, and naturally takes into account the above mentioned phenomena and theoretical challenges. We explicitly obtain the mean forces acting on the probe, as well as fluctuations (noise) from bath correlation functions evaluated in thermal equilibrium. While this scheme can be extended to higher orders, we stop here at the first nontrivial one, which is the second order. What is new or different from the major part of the literature is that our expansion around bath equilibrium is totally on the level of real-space trajectories and does not involve the Liouville equation or an expansion around the Fokker-Planck equation.  In that sense it deviates from the more usual approach as pioneered e.g. in Refs.~\cite{zwanzig,zwa61,zwan}.

Another aspect of this work is that it has before not been widely recognized that the friction and further terms in a nonlinear Langevin equation for a probe can be directly connected to the response behavior of the bath. It becomes thus apparent in the proposed scheme, that the nonlinear Langevin equation involves the time-symmetric bath observables (in contrast to linear order, where only time-antisymmetric observables, associated with the work done on the bath, enter). This makes explicit contact of the probe dynamics with what has been called the dynamical activity \cite{prl}, and gives an important insight: The nonlinear Langevin equation tests (or provides) much more dynamical details of the bath as compared to linear order, thus possibly opening the door for a wide range of novel phenomena. We hope  that the results exposed here are an important step in a statistical mechanical understanding of probe motion in visco-elastic media, and that the obtained equations contribute to the correct theoretical description for ongoing experiments exploring the dynamics of colloidal dynamics in visco-elastic fluids. Indeed a variety of new effects are being observed \cite{beh}, as visco-elastic fluids can easily be driven out of thermal equilibrium by a colloidal probe.

As a specific example, we discuss some properties of the genuine non-equilibrium state which is obtained by driving the probe with a constant velocity in a nonlinear bath.

The manuscript is organized as follows: We start with a description of the setup in section \ref{sec:set}.  Section \ref{epd2} is devoted to the physical aspects of the derivation and the required approximations. In Section \ref{2e} we give the new second order result. The more formal aspects of the computations regarding the use of response theory to derive the nonlinear Langevin equation is found in the Appendix \ref{rem2}. General properties are discussed in Section \ref{genp}. The example regarding a moving probe is  presented in Section \ref{exam}. 

\section{Effective Probe Dynamics}\label{epd}
\subsection{Setup}\label{sec:set}
We consider the problem of an object like a colloid or a small probe with position $x_t= (x_t(k),k=1,2,3)$, with spatial indices $k$, at time $t$ in contact with many degrees of freedom making up the medium. The degrees of freedom (constituting the configuration space) of the medium at time $t$ are abbreviated as $\eta_t$. The bath may itself already be represented in a reduced (e.g. mesoscopic) description. With the probe at rest, the bath is assumed in thermal equilibrium. The size of the probe (typically on the micron scale) is assumed large compared to the constituents of the bath (sizes below $\sim 100$~nm), so that the desired continuum description of the bath can be expected to prove useful.\\
The interaction potential between bath and probe is $U(x_t,\eta_t)$,  which, from the point of view of the bath, is a time-dependent external potential, depending on the instantaneous position of the probe $x_t$ as well as the (snapshot) configuration of the bath $\eta_t$.  We will consider that the rate of change of particle position, i.e.,  $\dot{x}_t$, is beyond the quasi-static regime, so that the bath is distorted by the probe's motion. The dynamics of the probe is thus influenced by the back reaction of the bath, for example exerting a friction force in response to motion. The force on the probe will naturally fluctuate as a function of time, which is captured in terms of a noise. \\
The rate of distortion (given e.g. through the velocity or acceleration of the probe) and the susceptibilities of the medium (its response behavior or relaxation time) are important parameters for the expansion. In case the distortion is large and/or the medium has large relaxation times, nonlinear response of the medium on the probe is relevant.\\ 
In the following, we derive the equation of motion of the probe particle by integrating out the dynamics of the bath, thereby demonstrating how the different orders in probe displacement appear. Mathematically, for the probe, we start from Newton's equation, valid at any instant in time for the position $x_t$ of the probe of mass $M$ in contact with the bath degrees of freedom $\eta_t$,
\begin{eqnarray}\label{new}
M\,\ddot x_t &=& -\nabla_x U(x_t,\eta_t) + f_t,
\end{eqnarray}
where in the following we use for the potential gradient the notation
\begin{eqnarray}
-\nabla_x U(x_t,\eta_t) &=:& F(x_t,\eta_t).\nonumber
\end{eqnarray}
We include an external force $f_t$, which is considered given and which will be irrelevant for the method of the paper. It could for example be due to an external potential for the probe, like gravity or the potential of an optical tweezer. 
\subsection{Physical Expansion Steps}\label{epd2}

There is no need to specify the dynamics of the bath, except that we assume throughout that the bath process $\eta_t$ is reversible when the probe is fixed at position $x$ (as it is then in equilibrium), with inverse temperature $\beta$. There is hence no arrow of time in $\eta_t$ when the probe is at rest. Furthermore, the bath has equilibrium free energy $\cal F(x)$.

The expansion will be performed around the case where the probe is at rest, i.e., we assume that previous positions $x_s$ with $s<t$, which contribute to the force at time $t$, are close to $x_t$. More specifically, it amounts to assuming that the work done by the probe on the bath within a time period of the bath relaxation time $\tau_\text{bath}$, by action-reaction estimated of the form $(x_{t-\tau_\text{bath}} - x_t)\cdot \nabla_x U(x_t,\eta_t)$, is small with respect to the temperature $\beta^{-1}$. (Or yet differently put, the time-integrated entropy flux per $k_B$ must remain small.)  Then, an expansion in $(x_s - x_t)$ is useful. A similar assumption will be imposed concerning the changes in the dynamical activity of the bath, for which we refer to later.

\subsection{Zeroth Order -- Quasi Static Regime}

The zero-th order approximation to the probe's dynamics consists in applying a purely adiabatic decoupling of the probe (the slow degrees of freedom) from the bath (the fast degrees of freedom).  It amounts to replacing 
\[
 F(x_t,\eta_t) \rightarrow \langle F(x_t,\eta_t) \rangle_{x_t}
\]
where the average $\langle\cdot\rangle_{x}$ is the equilibrium average of bath variables with the probe fixed at $x$.  Therefore,  $\langle \cdot\rangle_{x_t}$  is insensitive to previous probe positions and corresponds to the static limit for infinite time-scale separation between the (slow) probe and the (fast) bath degrees of freedom.  Moreover, the mean force
\[
\langle F(x_t,\eta) \rangle_{x_t} = -\nabla_x {\cal F}(x_t)
\]
 is the statistical force in thermal equilibrium at fixed probe position $x_t$, derived from the mentioned free energy $\cal F$ of the bath. For homogeneous systems, where the bath free energy does not depend on the probe position, it vanishes. In this limit, the probe dynamics \eqref{new} 
 is \begin{eqnarray}\label{ton}
  M\,\ddot x_t &=& -\nabla_x {\cal F}(x_t)+f_t,
 \end{eqnarray}
which is purely Newtonian and autonomous. No bath fluctuations and motion induced forces on the probe appear in this limit.
\subsection{First Order -- The Linear Langevin Equation}\label{fri}
In first order we add corrections due to the probe displacement to linear order in the displacements $x_s - x_t, s<t$. More specifically we split the force $F$ in \eqref{new} into a mean part and a fluctuating part
\begin{equation}\label{new1}
M\,\ddot x_t - f_t = -\nabla_x {\cal F}(x_t) + \left[\langle F(x_t,\eta_t) \rangle^{\omega^t}  +\nabla_x {\cal F}(x_t)\right] + \xi_t\\
\end{equation}
with
\begin{equation}\label{noi}
\xi_t := F(x_t,\eta_t)  - \langle F(x_t,\eta_t) \rangle^{\omega^t}.
\end{equation}
Here we have grouped the terms so that the correction to Eq.~\eqref{ton} becomes apparent (see the square bracket). 
  The average $\langle\cdot \rangle^{\omega^t}$ is over the bath $\eta_t$ which was evolved up to time $t$ under the given probe trajectory $\omega^t =(x_s, s<t)$. Naturally, the force acting on the probe at a given time depends on the history of probe positions. The noise $\xi_t$ constitutes the bath fluctuations.\\
There are two corrections with respect to \eqref{ton}.  \underline{First}, in square brackets, there is 
\begin{equation}\label{1co}
\langle F(x_t,\eta_t) \rangle^{\omega^t}  +\nabla_x {\cal F}(x_t)
\end{equation}
which is the mean force associated with probe motion, so that it is zero when the particle is at rest, i.e., if  $x_s=x_t, s<t$.  Treating it in first order perturbation theory for $x_s-x_t$, i.e.,  doing linear response and a short argument based on the Kubo formula exposed in Appendix \ref{rem2}, gives the well known friction force with memory,    
\begin{equation}\label{ff}
\langle F(x_t,\eta_t) \rangle^{\omega^t}  +\nabla_x {\cal F}(x_t) = - \int_0^{\infty} \id s \,\gamma_s(x_t) \,\dot{x}_{t-s}.
\end{equation}
The damping coefficient, or force memory kernel, is given by  (recalling spatial indices $i,k=1,2,3$)
\begin{equation}\label{nco1}
\gamma_s^{ki}(x_t) =  \beta\, \left\langle \frac{\partial U(x_t,\eta_0)}{\partial x(k)} \,;
\,\frac{\partial  U(x_t,\eta_s)}{\partial x(i)} \right\rangle_{x_t}.
\end{equation}
We introduced the covariance  $\langle A;B \rangle=\langle AB \rangle-\langle A\rangle\langle B \rangle$.  Let us be reminded that the average $\langle\cdot\rangle_{x}$ is the equilibrium average of bath variables with the probe fixed at $x$; $\gamma_s$ is thus a positive matrix.\\
For the \underline{second} correction we must tune the noise \eqref{noi} to be evaluated at zero order in probe displacement, replacing $\xi_t\rightarrow \xi^{(0)}$, and its covariance is of linear order in the displacements as well.  The correct noise to be used in this order of expansion is therefore
\begin{equation}\label{rep}
\xi_t^{(0)} = \nabla_x {\cal F}(x_t) +  F(x_t,\eta_t). 
\end{equation} 
It is then easy to verify the well known result that the friction matrix equals the force covariance, evaluated in equilibrium with the particle at rest at $x_t$ 
: in lowest order in probe displacement, we obtain the connection with the damping, 
\begin{align}\label{eq:n}
\gamma_s^{ki}(x_t) =  \beta\, \left\langle \xi_{s}^{(0)} (i);\xi_{0}^{(0)}(k)\right\rangle_{x_t} \approx \beta\, \left\langle \xi_{s}(i) ;\xi_{0}(k)\right\rangle^{\omega_t},
\end{align}
so that the noise covariance and the friction kernel have  the well known relation given by the fluctuation--dissipation relation.
In this linear order in probe displacement, the algorithm for simulating the probe motion thus follows the updating
\begin{eqnarray}\label{ab}
x_t &\longrightarrow & x_{t+\id t} = x_t + v_t \,\id t\\ 
v_t &\longrightarrow & v_{t+\id t} = v_t - \id t\,\int_0^{\infty} \id s \,\gamma_s(x_t) \,\dot{x}_{t-s} + \xi_t^{(0)}\,\id t\nonumber
\end{eqnarray}
where $\xi_t^{(0)} = \nabla_x {\cal F}(x_t) +  F(x_t,\eta_t)$ with the bath configuration  $\eta_t$ drawn at random from the equilibrium distribution $\langle \cdot\rangle_{x_t}$.  That is summarized in the stochastic differential equation
 \begin{eqnarray}\label{ton1}
  M\,\ddot x_t &=& -\nabla_x {\cal F}(x_t) - \int_0^{\infty} \id s \,\gamma_s(x_t) \,\dot{x}_{t-s} + \xi^{(0)}_t
 \end{eqnarray}
 which, together with the noise covariance in Eq.~\eqref{eq:n}, is well-known under the name of generalized Langevin dynamics, but note that here we did not specify the stochastic calculus for the noise $\xi_t$; \eqref{ton1} is just an abbreviation for \eqref{ab} and no further limits have been assumed. Because of the memory, this Langevin equation does already pick up certain features of visco-elasticity.  For example, when moving the probe at a fixed speed in some direction by an optical trap, it will typically go backwards upon switching off the laser if the memory is sufficiently strong, which is an aspect of elasticity.  Eq.~\eqref{ton1}  lacks however the influence of  nonlinear bath behavior.

It is interesting to note that this linear Langevin equation, based on Eq.~\eqref{ff}, is entropic, which we conclude from the fact that
\begin{equation}\label{wfor}
 \sum_i \int_0^{\infty} \id s \,\gamma_s^{ki}(x_t) \,\dot{x}_{t-s}(i) = \beta\,\left\langle  
W\,;\,\frac{\partial  U(x_t,\eta_t)}{\partial x(k)}\right\rangle^{\omega^t} 
\end{equation}
where
\begin{equation}\label{wee}
W = \sum_i\int_{-\infty}^{t} \frac{\partial U}{\partial x(i)} (x_s,\eta_s)\,\,\dot{x}_{s}(i)\,\id s
\end{equation}
is the work done on the bath from evolving under a time-dependent potential $U(x_s,\eta)$ as induced by moving the probe through it till time $t$. In other words, \eqref{nco1} is built from the correlation of the coupling force $F(x_t,\eta_t)$ with the entropy flux $\beta W$ per $k_B$ up to time $t$, a purely dissipative contribution.  

We close this subsection by discussing a few limiting cases. For homogeneous systems, the average in Eq.~\eqref{nco1} does not depend on $x_t$. If moreover the coupling potential between the probe and the medium is linear, i.e., if $\partial_{x_k} U(x,\eta) = a_k(\eta)$, then
\begin{equation}\label{appa}
\gamma_s^{ki}(x_t) = \gamma_s = \beta\,\left\langle a_k(\eta_0)\,;
\,a_i(\eta_s)\right\rangle
\end{equation}
over the medium equilibrium. At any rate, the damping $\gamma_s$ is $O(\lambda^2)$ in the coupling strength $\frac{\partial U(x,\eta)}{\partial x(k)} = O(\lambda)$ and is negligible for times $s$ beyond the bath relaxation time $\tau_\text{bath}$. In a Markovian approximation  the friction force \eqref{ff} is simplified,
to
\begin{align}
\int_{0}^\infty \id s\, \gamma_s(x_t)\,{\dot x}_{t-s}\approx \dot x_t \int_{0}^\infty \id s\, \gamma_s(x_t). 
\end{align}

\subsection{Second Order -- Nonlinear Langevin Equation}\label{2e}
Coming to the correction to Eq.~\eqref{ton}, in second order in probe displacement, we finally  arrive at the main aim of this paper. We apply second order perturbation theory to \eqref{1co}.

There are in general two contributions to second order, the first being due to nonlinear coupling in the bath probe potential $U(x_t,\eta_t)$ in Eq.~\eqref{new} (where temporal bath-correlations between $\partial_x F(x,\eta_0)$ and $F(x,\eta_s)$ would enter). We will in the main text discard these, and concentrate on  what we believe is the physically more interesting case where  the coupling is linear, i.e., $\partial_{x_k} U(x,\eta) = a_k(\eta)$ as in \eqref{appa}. The remaining second order effect lies then in the bath dynamics itself. 

Specifically, one has to add to the right-hand side of \eqref{ff} the force term which is quadratic in $x_s-x_t$, and analogously move to first order in the noise-replacement \eqref{rep}.\\ 
The second order contribution to the mean force in the Langevin dynamics requires an understanding of how the bath responds in second order perturbation theory to the probe's motion.  
Beyond the merely formal second order Taylor expansion, this means a departure from the usual territory of purely dissipative effects and involves what has been called the (change in) {\it dynamical activity} of the bath dynamics, $\cal{D}$.  The latter is 
 identified as the time-symmetric part in the action for describing the modified probability weights on the bath path-space due to the probe displacement. See e.g. \cite{prl,ub} where it is also called the frenetic contribution to the response. The underlying heuristics is that a perturbation in the bath can also influence the bath kinetics in its time-symmetric sector, {\it like when the trading volume in a financial market is also influenced, and not only the interest rates, when an important financial player enters or opportunities arise}.\\
   Before expressing the second order in terms of $\cal{D}$, we aim to discuss some of its properties; 
  $\cal{D}$ can be obtained in practice from linear response experiments for {\it time-symmetric} observables. In particular, when disturbing the bath by moving the probe in the $j$-direction, the \emph{linear} response of a time--symmetric observable $O_\text{sym}$ depending on the bath trajectory up to time $t$ is given by
 \begin{equation}\label{lres}
 \langle O_\text{sym} \rangle^{\omega^t} - \langle O_\text{sym} \rangle_{x_t} = \frac{\beta}{2}\,\int_{-\infty}^t\id s \,(x_t(j)-x_s(j)) \,\langle {\cal D}^{x_t}_j(\eta_s)\,O_\text{sym}\rangle_{x_t}.
 \end{equation}
 Time--symmetric observables include for example a momentum current or an even moment of particle currents in the bath. 
 As a simple case when the bath is overdamped and can be characterized by a backward generator $L_{x}$ when the probe is at position $x$ (i.e., the backward Smoluchowski operator), the dynamical activity for our  case reads ${\cal D}_j^{x_t}(\eta_{v}) =  L_{x_t}\frac{\partial U}{\partial x(j)}(x_{t},\eta_{v})$, as can be read from formula (7) in \cite{ub}. 
 We can now apply this to identify the correction to the Langevin equation in quadratic order. We add one more term to Eq.~\eqref{ff} (again, using already $\partial_{x_k} U(x,\eta) = a_k(\eta)$)
 \begin{align}\label{ff2}
&\frac{\partial}{\partial x(k)} {\cal F}(x_t) - \langle a_k(\eta_t) \rangle^{\omega^t}  =  - \beta\sum_i\int_0^\infty  \dot x_{t-s}(i)\,\,\left\langle a_i (\eta_0) \,a_k (\eta_s)\right\rangle_{x_t}\,\id s \\
& - \frac{\beta^2}{2}\sum_{ij} \int_0^\infty\id s\int_0^\infty \id s' \dot{x}_{t-s}(i)\,\left(x_t(j) - x_{t-s'}(j) \right)\, \left\langle {\cal D}_j^{x_t}(\eta_{s})
  a_i (\eta_{s'})\, a_k (\eta_{s+s'})\right\rangle_{x_t}\nonumber.
 \end{align}
Note again that we have omitted terms due to nonlinear bath-probe coupling. The last line is as we believe entirely novel and adds another (besides memory) elastic component to the equation. It correlates the product of coupling force and entropy flux with the linear response kernel for time-symmetric bath observables (i.e., to ${\cal D}$). The resulting three-time correlation function in the equilibrium bath, especially including the change in dynamical activity, is expected to add a new time-scale to the system which is not purely dissipative. We hope that such correlation functions can in principle be measured independently, for example via scattering experiments  (cf. \cite{vanh,ven} for the traditional theory). 
  $\cal{D}$ depends on bath details, and  there are various specific mathematical forms of ${\cal D}$ for different baths, see the examples in \cite{ub}. 
 Note also the prefactor $\beta^2$.

We are now able to state the nonlinear (to quadratic order) Langevin equation.  The probe dynamics satisfies, in  quadratic order in its displacements, 
\begin{align}\label{resu}
&M\ddot x_t(k) - f_{t}(k) = -\partial_k{\cal F}(x_t)- \beta\sum_i\int_0^\infty  \dot x_{t-s}(i)\,\,\left\langle a_i (\eta_0) \,a_k (\eta_s)\right\rangle_{x_t}\,\id s \\
& - \frac{\beta^2}{2}\sum_{ij} \int_0^\infty\id s\int_0^\infty \id s' \dot{x}_{t-s}(i)\,\left(x_t(j) - x_{t-s'}(j) \right)\, \left\langle {\cal D}_j^{x_t}(\eta_{s})
  a_i (\eta_{s'})\, a_k (\eta_{s+s'})\right\rangle_{x_t}\nonumber+\xi_t^{(1)}(k).
\end{align}
The  noise, again, exactly given by $\xi_t = F(x_t,\eta_t) - \langle F(x_t,\eta_t) \rangle^{\omega^t}$ should now be treated in first order approximation in \eqref{fri}, similarly to  $\xi^{(0)}$ in Eq.~\eqref{rep} being the zero-th order approximation. So we define in this order,
\begin{equation}\label{rep1}
\xi_t^{(1)} = 
\nabla_x {\cal F}(x_t) + \int_{-\infty}^t \id s \,\gamma_s(x_t) \,\dot{x}_{t-s} + F(x_t,\eta_t)
\end{equation}
which adds one term relative to Eq.~\eqref{rep}. Its variance is then taken under the average which is linearly perturbed by $\langle\cdot\rangle_{x_t}$. (The latter may be identified  with the so called McLennan distribution, giving the first order correction to the bath equilibrium distribution from having small displacements $x_s-x_t$, see \cite{mac}).  
The covariance of the noise in \eqref{resu} is then  
\begin{eqnarray}
\langle \xi_{t}^{(1)}(k)\,;\,\xi_{s}^{(1)}(i)\rangle &=&
\langle a_k(\eta_t)\,;\,a_i(\eta_s)\rangle_{x_t} - \frac{\beta}{2}\,\langle (\tilde{\cal{D}}+W)\,\,a_k(\eta_t)\,a_i(\eta_s)\rangle_{x_t} \nonumber\\
&+&\frac{\beta}{2} [\langle W a_k(\eta_t)\rangle_x\langle a_i(\eta_s)\rangle_{x_t} + \langle W a_i(\eta_s)\rangle_{x_t}\langle a_k(\eta_t)\rangle_{x_t}] 
\label{efn}
\end{eqnarray}
where we have used
$\beta W$ the entropy flux in Eq.~\eqref{wee} and introduced the abbreviation
\begin{align}
\tilde{\cal{D}}=\int_{-\infty}^t \id s'\sum_{i} \left(x_t(i) - x_{t-s'}(i) \right)\,  {\cal D}_i^{x_t}(\eta_{s'})
\end{align}
for the dynamical activity as defined before.

Eqs.~\eqref{resu} and \eqref{efn} constitute our main results, which allow to compute the stochastic trajectory to second order in probe motion. Note that there is now more than just memory to evoke the elastic properties of the bath.  There also appears another time-scale.
For visco-elastic fluids, there is first the time-scale of dissipation through which the absorbed energy delivered by the probe is released in internal degrees of freedom of the medium or in the environment.  Secondly, there is an elastic time-scale which refers to time-symmetric response of the fluid, which appears more of kinetic nature and relates to changes in the dynamical activity of the fluid induced by the probe's motion and appears in the three point correlation function.

\section{Symmetries}\label{genp}
In the above expansion we have stopped at second order (in probe displacements or velocity). Certain symmetries imply that the extra term of quadratic order is  sometimes less relevant.
Generally, we note that for systems which are isotropic and homogeneous, the quadratic order must vanish identically.  The second order is thus important for sufficiently non-symmetrically shaped probes  or inhomogeneous or anisotropic baths.\\ Inhomogeneities could for example arise near surfaces or boundaries, or in external fields like gravity, where density gradients in the bath are present. In general, the mean velocity of the probe can be nonzero (for example a probe drifting towards less dense regions, or being attracted/repelled by surfaces due to (Casimir) forces mediated by the bath), even in the absence of external forces $f$ in Eq.~\eqref{resu}. In such situations, a Markovian limit can exist but the probe dynamics is generally not in equilibrium, even in the absence of external forces $f$, as the particle is drifting.   One may easily convince oneself that for a homogeneous system, the mean squared displacement as well as the linear response mobility of the probe are unchanged due to symmetries. This is expected to change when including a third order correction (see e.g. Refs. \cite{Khandekar15, Soo16}).\\
 Anisotropic baths could be given by complex baths undergoing ordering transitions, such as e.g. liquid crystals  or ferrofluids. 
 If the bath is homogeneous, but anisotropic, the situation is different from the above. Here (always assuming that the bath itself is in equilibrium) due to translational invariance, one must have $\langle \dot x\rangle=0$ in the absence of external forces. Because if this, and taking the mean of Eq.~\eqref{resu}, the Markovian limit of the second order term must vanish. 
 For these baths, regarding the case without external forces, the probe dynamics is in equilibrium.  It is however interesting to consider the probe's motion itself as an anisotropic perturbation, when moving it in one direction as we do next. If the bath is anisotropic, this induces important effects through second order.

\section{Example: Steadily Driven Probe}\label{exam}
Consider a probe (for simplicity of notation in two dimensions) in a potential $V(x(1),x(2))=\frac{1}{2}[\kappa x(2)^2+\tilde\kappa (x(1)-v_1t)^2]$, e.g. realized experimentally by use of an optical tweezer. If  $\tilde\kappa\to\infty$,  the probe moves  with a constant velocity $v_1$ in direction $1$, and fluctuates in a quadratic potential in direction 2. We assume the bath is homogeneous. The probe's displacement in direction 2 is small and is well described by the linear equation like \eqref{ton1} but with an effective friction and noise, 
\begin{align}\label{eq:example}
&M\ddot x_t(2) + \kappa x_t(2) = - \beta\int_0^\infty  \dot x_{t-s}(2)\,\gamma^{\textrm{eff}}(s) +\xi_{t}^{\textrm{eff}}(2).
\end{align} 
The effective friction kernel is affine in $v_1 $ for sufficiently small $v_1$:
\begin{align}
\nonumber&\gamma^{\textrm{eff}}(s)=\left\langle a_2 (\eta_0) \,a_2 (\eta_s)\right\rangle 
+ v_1\frac{\beta}{2} \int_0^\infty\id s'  \, \int_{-\infty}^{s'} \id\nu\left\langle {\cal D}_2^{x_t}(\eta_{s}) a_1(\eta_{\nu})\, a_2 (\eta_{s+\nu})\right\rangle\\
&+ v_1\frac{\beta}{2} \int_0^\infty \id s' \,s'\left\langle {\cal D}_1^{x_t}(\eta_{s}) a_2 (\eta_{s'})\, a_2 (\eta_{s+s'})\right\rangle.
\end{align}
The noise correlator is also dependent on $v_1$, and reads for sufficiently small $v_1$ 
(we assume that $\langle a_2 (\eta_{s})\rangle=0$ for the homogeneous system)
\begin{align}
\nonumber\langle \xi_{t}^{\textrm{eff}}(2)\,;\,\xi_{s}^{\textrm{eff}}(2)\rangle =
&\langle a_2(\eta_t)\,;\,a_2(\eta_s)\rangle_{x_t} -\frac{\beta}{2}\,v_1 \int_{-\infty}^t \id s' \, s'\,\left\langle {\cal D}_1^{x_t}(\eta_{s}) a_2 (\eta_{t})\, a_2 (\eta_{s})\right\rangle\\
&+ \frac{\beta}{2}\,v_1\int_{-\infty}^t \id s'  \,\left\langle a_1(\eta_{s'}) a_2 (\eta_{t})\, a_2 (\eta_{s})\right\rangle.
\label{efne}
\end{align}
Because of the last term it is clear that the Einstein relation (or the fluctuation--dissipation theorem) is not valid for the $2-$direction, so that the non-equilibrium nature of the dynamics of the 2-component is evident.
\section{Conclusion}
We have derived the generalized nonlinear Langevin equation that describes the dynamics of a probe in visco-elastic or effectively nonlinear media.  There appears a non-dissipative contribution  which marks the contrast with motion in Newtonian fluids.  In turn that contribution enters another time-scale in the probe's motion which is related to the time-symmetric fluctuations  of the medium, visible in second order response around its thermal equilibrium. The presented Langevin description for colloidal dynamics in visco-elastic fluids (and the scheme for inclusion of higher orders) opens the way for systematic understanding of various observed phenomena (nonlinear rheological properties) as will be the subject of follow-up papers.
\begin{acknowledgments}
We thank C. Bechinger, J. R. Gomez-Solano and H. Soo for useful discussions. M.K. was supported by Deutsche Forschungsgemeinschaft (DFG) grant No. KR 3844/2-1.
\end{acknowledgments}

\begin{appendix}
\section{Nature of the Expansion}\label{rem2}
The evolution equation \eqref{resu} appears from estimating the response of the visco-elastic thermal bath to the probe's displacements.  This naturally involves the response to second order, expanded around equilibrium (in contrast to Refs.~\cite{m1,m2}, where the baths are per se out of equilibrium, but otherwise a similar strategy was used).

To start, there is the reference condition with expectations $\langle\cdot\rangle_{x_t}$ which is the one of thermal equilibrium of the bath at fixed probe position $x_t$.  The reference process for the bath is thus with a potential constant  in time, equal to  $U(x_t,\eta)$ (the one at time $t$).  On the other hand, the real situation is that the probe moves which can be viewed as a perturbed situation for the bath-dynamics.  We assume that the perturbed process started from thermal equilibrium at time zero with probe position $x_{0}$ and then the bath evolves with time-dependent potential
\begin{eqnarray}\label{2st}
U(x_s,\eta) &=& U(x_t,\eta) - V_s(\eta), \quad s\leq t\\
V_s(\eta) &=&  (x_s - x_t)\cdot F(x_t,\eta) - \frac 1{2}\sum_{i,j}(x_s(i) - x_t(i))(x_s(j) - x_t(j))\,\partial^2_{ij}U(x_t,\eta) +\ldots\nonumber
\end{eqnarray}
where the second derivatives $\partial^2_{ij}$ are with respect to the probe position coordinates $x_i,x_j$.  That second order term of course vanishes in the case of a coupling which is linear in the probe position -- as considered in the main text -- and where
\begin{equation}\label{1st}
U(x,\eta) = x\cdot a(\eta), \qquad V_s(\eta) = - (x_s - x_t)\cdot a(\eta)
\end{equation}
is  true.  For simplicity of the set-up we restrict ourselves here to the case \eqref{1st}, in which we concentrate rather on the nature of second order response theory.  It is here mathematically useful to extend the probe's history to the far past where we put $x_{s=-\infty} = x_t$; that has no physical influence.

We recall the main result of \cite{ub} to describe the second order response around equilibrium for time-dependent potential perturbations as adapted to the present situation. The observable in question, whose expectation under $\langle \cdot \rangle^{\omega^t}$ must be evaluated, is $a(\eta_t)$.  We consider thus the difference 
\[
 \langle a(\eta_t)\rangle^{\omega^t} - \langle a(\eta)\rangle_{x_t}
 \]
 (All averages are with respect to the bath process; $\langle \cdot\rangle^{\omega^t}$ is the perturbed process with potential $x_s\cdot a(\eta)$ at time $s<t$ and started long ago from drawing $\eta_{-\infty}$ at thermal equilibrium with probe at position $x_t$, and $\langle \cdot\rangle_{x_t}$ is the reference equilibrium process with fixed potential $x_t\cdot a(\eta)$ and also started from drawing $\eta_{-\infty}$ at thermal equilibrium with probe at position $x_t$.)  We use equation (5) in \cite{ub},
 \begin{equation}\label{linor}
 \langle a(\eta_t)\rangle^{\omega^t} - \langle a(\eta)\rangle_{x_t} =  \langle S\,;\,a(\eta_t)\rangle_{x_t} - \langle {\cal D}\,S\,a(\eta_t)\rangle_{x_t}
 \end{equation}
 where we need to explain the meaning of $S$ and $\cal D$.  First about $S$; it has a thermodynamic meaning as  the generated entropy flux per $k_B$ over $[0,t]$: from the First Law,
 \begin{eqnarray}\label{wok}
 S &=& \beta\left[V_t(\eta_t) -V_{-\infty}(\eta_{-\infty})  -\beta\,\int_{-\infty}^t\id s \,\frac{\partial V_s}{\partial s}(\eta_s)\right]\nonumber\\
 &=&   \beta\,\int_{-\infty}^t\id s \,\dot{x}_s\cdot a(\eta_s)
 \end{eqnarray}
 As a consequence, the linear order in \eqref{linor} gives
 \begin{equation}\label{linor1}
  \langle a(\eta_t)\rangle^{\omega^t} - \langle a(\eta)\rangle_{x_t} =   \beta\,\int_{-\infty}^t\id s \,\langle\dot{x}_s\cdot  a(\eta_s)\,;\,a(\eta_t\rangle_{x_t} 
  \end{equation}
  which suffices for establishing the (familiar linear generalized) Langevin equation \eqref{ton1}, or
 \begin{equation}\label{arr1}
 M\ddot x_t - f_t= -\nabla_x {\cal F}(x_t) - \int_0^\infty \gamma_s(x_t) \cdot \dot x_{t-s}\,\id s +
 \xi_t^{(0)}(x_t)
 \end{equation}
 with
 \begin{equation}\label{eir}
 \gamma_s^{ki}(x) = \beta\,\,\langle \xi_{s}^{(0)}(k)\,;\xi_{0}^{(0)}(i) \rangle,\quad s>0.
 \end{equation}
The time-scale of the friction and the noise is dictated by the force-force time-correlation. The equilibrium condition of the $\eta-$bath has given rise to (1) the systematic (mean) force being derived from the free energy $\cal F$, and (2) the Einstein relation \eqref{eir} between friction and noise. In fact we have shown here how the {\it first} fluctuation--dissipation relation (say, the Kubo formula \cite{kubo,tod92}) yields the {\it second} fluctuation--dissipation relation (say, the Einstein formula). Additional external forces $f_t$ can be added to \eqref{arr1}, even nonconservative ones, as long as the displacements of the probe on the (fast) time-scale of the bath can be treated in linear order response.  Then when driving with $f_t$, the Einstein relation \eqref{eir} remains valid but under additional non-equilibrium driving the Sutherland-Einstein relation between diffusion constant and mobility can be and in general will be violated; see e.g. \cite{Blickle07,Lander11, Kruger11,royal1,royal2}.
 
  We have used here that the work done by the probe on the bath is small compared to $k_BT$ but also that we can neglect  the last term in \eqref{linor}.  What we want presently is to go to second order around equilibrium.  That becomes relevant for higher speeds or when the medium is more susceptible to time-dependent perturbations.  That is the case for visco-elastic media which have much slower relaxation times than Newtonian fluids.
To truly deal with visco-elastic effects, we need, as a first step (with respect to even higher orders in displacement), to include therefore the last term in \eqref{linor} and that is the subject of the next section.

The next step is to go one more order from \eqref{arr1}.  The first and somewhat trivial change is to use \eqref{2st} up to quadratic order, i.e., no longer assuming that the coupling with the bath is linear in the probe position.  That amounts to adding an extra term to the last line of \eqref{wok}
and it changes \eqref{linor1} into
\begin{eqnarray}\label{linor2}
  \langle F(x_t,\eta_t)\rangle^{\omega^t} - \langle F(x_t,\eta)\rangle_{x_t} &=&  - \beta\,\int_{-\infty}^t\id s \,\langle\dot{x}_s\cdot  F(x_t,\eta_s)\,;\,F(x_t,\eta_t\rangle_{x_t}\\ -&\beta& \sum_{ij}\int_0^t\id s\,\dot{x}_s(i)\,(x_s(j)-x_t(j))
  \, \left\langle \partial_{ij}^2U(x_t,\eta_s) \,;\,F(x_t,\eta_t)\right\rangle_{x_t}\nonumber
  \end{eqnarray}
That addition vanishes for a linear coupling as used in the main text.  More interesting is to explore the physics of the second term in \eqref{linor}.\\
As announced around equation \eqref{lres} the ${\cal D}$ in \eqref{linor} governs the linear response around equilibrium for time-symmetric observables.  It contains non-thermodynamic information about the bath and takes the following general form
\begin{equation}\label{dd}
{\cal D} = \frac{\beta}{2}\int_{-\infty}^t (x_s(j) - x_t(j))\,{\cal D}_j^{x_t}(x_t,\eta_s)\,\id s.
\end{equation}
Through ${\cal D}_j^{x_t}$ (which is a force component per unit time) the probe can possibly feel the difference between baths which are thermodynamically identical but still kinetically different. That exactly fits with the idea of visco-elastic media where it is for example not only the temperature or the density of the bath which plays a role when moving around but also viscosities that tell how the momentum current in the bath responds to shearing.\\
\end{appendix}

\end{document}